# K$_2$RENb$_5$O$_{15}$ (RE=Ce,Pr,Nd,Sm,Gd-Ho)：A family of Quasi-One-Dimensional spin-chain compounds with large interchain distance

Qingyuan Zeng,[a,‡] Han Ge,[b,†] Maofeng Wu,[a] Shaoheng Ruan,[a] Tiantian Li,[b] Zhaosheng Wang,[c] Jingxin Li,[c] Langsheng Ling,[c] Wei Tong,[c] Shuai Huang,[d] Andi Liu,[a] Jin Zhou,[a] Zhengcai Xia,[a] Jieming Sheng,[b,e] Liusuo Wu,[b,e*] Zhaoming Tian[a*]

[a] Wuhan National High Magnetic Field center and School of Physics, Huazhong University of Science and Technology, Wuhan, 430074, China.

[b] Department of Physics, Southern University of Science and Technology, Shenzhen, 518055, China

[c] Anhui Key Laboratory of Low-energy Quantum Materials and Devices, High Magnetic Field Laboratory, HFIPS, Chinese Academy of Sciences, Hefei, Anhui 230031, China

[d] Key Laboratory of Novel Materials for Sensor of Zhejiang Province, Institute of Material Physics, Hangzhou Dianzi University, Hangzhou 310018, China

[e] Shenzhen Key Laboratory of Advanced Quantum Functional Materials and Devices, Southern University of Science and Technology, Shenzhen, 518055, China

**ABSTRACT:**

One-dimensional (1D) spin chain systems have received special attention to discover the novel magnetic ground states and emergent phenomena, while the magnetic studies on rare-earth (RE)-based 1D spin chain materials are still rare. Here, we report the synthesis, structure and magnetic behaviors on a family of tetragonal tungsten-bronze (TTB) structure K$_2$RENb$_5$O$_{15}$ (RE = Ce, Pr, Nd, Sm, Gd-Ho) compounds, which consist of 1D linear spin-chain structure built by RE$^{3+}$ ions along the *c*-axis and well spatially separated by the nonmagnetic K/Nb-O polyhedrons with large interchain distances of ~ 8.80-8.88 Å in the *ab*-plane. The low temperature magnetic measurements reveal the absence of long-range magnetic order down to 1.8 K for all serial K$_2$RENb$_5$O$_{15}$ compounds and the dominant ferromagnetic interactions for RE=Ce,Dy and antiferromagnetic interactions for other members. Among them, K$_2$GdNb$_5$O$_{15}$ with spin only magnetic moment *S*=7/2, exhibits a long-range magnetic order with $T_N$~0.31 K and strong spin fluctuations at low temperatures due to its low-dimension characteristics. Moreover, a large magnetocaloric effect under low field change (∆*B*) of ∆*B* = 0-2 T is realized at temperatures below 1 K for K$_2$GdNb$_5$O$_{15}$, letting it as an ideal candidate for adiabatic magnetic refrigeration applications at sub-kelvin temperatures. The K$_2$RENb$_5$O$_{15}$ become a rare family of insulting RE-based magnets to explore the novel 1D spin chain physics beyond the 3*d* TM-based counterparts, in terms of its combination of low dimension, strong spin-orbital coupling and the rich diversity of RE ions.

---

[‡]Qingyuan Zeng and Han Ge contribute equally to this work.

# INTRODUCTION

One-dimensional (1D) quantum spin chain (SC) materials as important paradigm of low-dimensional magnets, have attracted much attention as the platform to discover the exotic magnetic states predicted by theoretical models as well as the related quantum phenomena.[1-4] Compared to the high dimensional frustrated magnets, 1D interacting spin systems can host the exotic magnetic ground states without geometric frustration owing to the intrinsic low dimensionality and low coordination number. In this system, a variety of non-trivial quantum phenomena such as the Tomonaga-Luttinger (TL) liquid,[5,6] ferrotoroidic order,[7] Bose-Einstein condensation and quantum criticality with emergent excitations[8-10] have been theoretically proposed and experimentally reported, and also a large variety of materials with quasi-1D structural motif have been investigated including the uniform and alternative linear spin chain, zigzag spin chain, spin ladder and 1D spin tubes.[3,10-13] Among them, the $S$=1/2 Heisenberg antiferromagnetic linear spin chain (AFMC) with only nearest-neighbor (NN) interactions have an exactly solvable ground that is macroscopically entangled TL liquid state exhibiting topological spinon excitations.[6,13] To experimentally realize this exotic state, a large number of compounds especially those containing the 3d transition-metal (TM) ions have been discovered and characterized in detail in past several decades.[10,13-15] While, the real materials always undergo a three-dimensional (3D) long-range magnetic order due to the non-negligible interchain exchange interactions or structural distortions, then the genuine 1D magnetic phase and related exotic magnetic phenomena remains not be fully understood.

Recently, the quasi-1D spin systems containing 4f rare-earth (RE) ions have also garnered intensive attention. Compared to the 3d TM based systems, the intrinsic strong spin-orbit coupling (SOC) and crystalline electric field (CEF) associated with 4f electrons of RE ions bring the highly anisotropic magnetic exchange interactions,[16,17] which let RE-based spin chain systems to possibly realize the novel magnetic phases. Moreover, the exchange interactions between $RE^{3+}$ local moments are relatively small due to its more localized nature of 4f electrons, which allow an easy tunability of spin states through the external field accessible in the laboratory. Until now, the studied quasi-1D RE-based materials can be divided into two categories:[18-21] one is the intermetallic compound containing conductive electron, the other is the insulating oxide with magnetism fully determined by the exchange interactions of 4f local moments. For the former, owing to the hybridization between conductive electrons and 4f local moments, the Ruderman-Kittel-Kasuya-Yosida (RKKY) interaction and Kondo effect between the local moments complicate the underlying physics of exotic magnetic behaviors.[18,19] By contrast, the latter is attractive to unveil the intrinsic 1D spin chain physics due to the free of conductive electrons but is less studied, only a few RE-based spin chain systems have been magnetically characterized. The formates RE(HCOO)$_3$ (RE=Gd-Er) is a relatively well-studied systems where the RE ions form a uniform 1D spin-chain structure along the c-axis and triangular-lattice topology perpendicular to the chain direction,[20,21] and long-range magnetic order is developed due to the comparable interchain RE-RE distance ($d_{inter}$ ~6.5-6.8 Å) with intrachain separation ($d_{intra}$ ~3.9-4.1 Å). The very recently discovered calcium oxyborates $Ca_4REO(BO_3)_3$ (RE=La-Lu) provide a 1D spin-chain with well separated interchain distance $d_{inter}$~8.1-9.2 Å compared to $d_{intra}$ ~3.6-3.8 Å,[22] while the antisite disorder between $RE^{3+}$ and $Ca^{2+}$ cations also induce the magnetic disorder effect as an obstacle on unveiling its intrinsic magnetic ground state.[23] Therefore, the exploration of new materials possessing both large RE-RE interchain distances and weak chemical disorder are essential to study the pure 1D exotic quantum physics.

From the view of application, low dimensional RE-based magnets are attractive as promising magnetic refrigeration materials at low temperatures. The large magnetocaloric (MC) effect can be used to obtain the sub-Kelvin (< 1 K) temperatures during the adiabatic demagnetization (ADR) process, due to the reduced transition temperature and highly spin fluctuated magnetic ground state induced by its low dimension nature.[24-26] This ADR technology is important for sub-Kelvin space applications under microgravity conditions and quantum computing.[27,28] Usually, the lowest achievable temperature for magnetic cooling is related to the magnetic ordering temperature of materials, thus the nearly non-interacting dilute paramagnetic salts such as cerous magnesium nitrate (CMN) and ferric ammonium (FAA) are widely used in ADR as refrigerant.[29,30] In the case of quasi-1D spin-chain systems, the disorder or magnetic order state at suppressed temperatures induced by low dimensionality let them as superior refrigerant to achieve ultra-low temperature,[31,32] and the large residual magnetic entropy around the transition point can produce sizeable MC effect. For the RE-based materials, large cooling capacity, weak exchange interaction and diverse magnetic anisotropy make them flexible for designing MC materials working at low temperatures and different field regions. However, the experimental identification on MC effect in the RE-based 1D spin chain remains a nearly unexplored research topic, the related experimental studies on ideal 1D materials are highly desirable.

The RE-based tetragonal tungsten-bronze (TTB) structure $K_2RENb_5O_{15}$ (RE=La-Ho) compounds as an important family of ferroelectric materials, have been well studied for their interesting dielectric and nonlinear optic properties.[33-35] The previous crystal structure and temperature (T) -dependent dielectric measurements reveal that most members of $K_2RENb_5O_{15}$ (RE=La-Ho) compounds undergo two phase transitions from the paraelectric (PE) to antiferroelectric (AFE) to ferroelectric (FE) phases as function of temperatures.[33] Correspondingly, they crystallize into the centrosymmetric space group of P/4mbm (no. 127) and polar P4bm (no. 100) at the high and low temperature regimes, and the AFM-FE transition temperature ($T_C$) ranges from ~300 K to ~460 K depending on the choice of $RE^{3+}$ ions.[34,35] Despite the well-studied crystal structure and related electronic properties, the lattice geometry of $RE^{3+}$ ions and their magnetic behaviors have not been unveiled, motivating the present study. Here, we report the synthesis and magnetic properties on this serial RE-based $K_2RENb_5O_{15}$ (RE=Ce-Ho) oxoniobates, in which the magnetic $RE^{3+}$ ions form a 1D uniform chain along the *c*-axis with intrachain distance of $c$~3.89-3.92 Å and well spatially separation by the $NbO_6$ octahedron with large interchain distance of $a/\sqrt{2}$ ~ 8.79-8.87 Å in the *ab*-plane. The square-lattice arrangements of the infinite columns of $RE^{3+}$ spin chains are distinct from the ones of $RE(HCOO)_3$ and $Ca_4REO(BO_3)_3$ with triangular lattice topology. The magnetic characterization and electron spin resonance (ESR) spectra reveal the different types of magnetic interactions and magnetic anisotropy of $K_2RENb_5O_{15}$ in respect to different RE ions. Moreover, $K_2GdNb_5O_{15}$ shows a long-range magnetic order below $T_N$~0.31 K and exhibits large magnetic entropy change $\triangle S_m$~10.8 J/K/(mol-Gd)=0.62Rln(2S+1) at low-field change ($\Delta B$) of 0-2 T and temperature $T$~ 1 K, as a suitable material for magnetic refrigeration in sub-Kelvin temperature regimes with good magnetocaloric performance.

## ■ EXPERIMENTAL SECTION

**Solid-state Synthesis.** The $K_2RENb_5O_{15}$ (RE =La, Ce,Pr, Nd, Sm-Ho) polycrystalline samples were synthesized by a solid state reaction method using $K_2CO_3$(99.99%), $Nb_2O_5$(99.5%) and RE oxides (RE=La, Ce, Pr, Nd, Sm-Ho; 99.99%) as starting materials. Before using, raw materials $RE_2O_3$ (RE=Pr, Nd, Sm) were pre-dried at 800°C for 12 hours in air. Stoichiometric amounts of starting materials with mole ratio of K:RE:Nb=2:1:5 were weighted, and ground for several hours. After that, the mixtures were pre-reacted in air at temperature 1100°C for 24 hours. For $K_2RENb_5O_{15}$ (RE=La, Ce, Pr, Nd, Sm) samples, the products were re-ground and reacted at temperature of 1150 °C-1200 °C for 4 days with intermediate grindings. For RE=Gd-Ho compounds, higher reaction temperature of 1250 °C is used to obtain pure phase samples.

**Structural determination.** The phase purity and crystal structure of $K_2RE_2Nb_5O_{15}$ were checked by room temperature powder X-ray diffraction (XRD, Rigaku Smartlab) with Cu Kα radiation (λ= 1.5418 Å). The Rietveld refinements of XRD data were performed for structural analysis using Material Studio software.[36,37] Quantitative calorimetric analyses by differential scanning calorimetry (DSC) were carried out on the polycrystalline samples at a scanning rate of 10K/min using a Perkin–Elmer Pyris DSC Diamond under the argon atmosphere, which can be used to determine the structural transition as function of temperature.

**Physical property measurements.** The dielectric properties were measured by a precise impedance analyzer (Model No: Wayne Kerr 6500B), the high-temperature parts were performed with the Partulab DMS-2000 dielectric measurement system (Partulab Technology Co, China) and low-temperature ones were connected with the Physical Property Measurement System(PPMS,Quantum Design). The magnetic properties were measured using a commercial SQUID magnetometer (MPMS, Quantum Design) and commercial PPMS System equipped with a vibrating sample magnetometer (VSM) option. The pulsed field magnetizations up to 50 T were measured by the induction method at Wuhan National High Magnetic Field Centre (WHMFC) with a calibration by the DC magnetization data. The X-band (9.4 GHz) electron spin resonance (ESR) measurements were carried out using a Bruker spectrometer at the High Magnetic Field Laboratory of the Chinese Academy of Science. The specific heat $C_p(T)$ measurements of $K_2GdNb_5O_{15}$ were performed using a Quantum Design physical properties measurement system (PPMS) in the temperatures above 2 K, and at lower temperatures down to 0.08 K, the specific heat was measured using the PPMS equipped with a dilution refrigeration by the heat capacity options.

**Table 1.** The selected bond distances, bond angles and RE−RE distances of $K_2RENb_5O_{15}$ (RE= Ce—Ho) polycrystals

| RE | Ce | Pr | Nd | Sm | Gd | Tb | Dy | Ho |
|---|---|---|---|---|---|---|---|---|
| Crystal system | Tetragonal | Tetragonal | Tetragonal | Tetragonal | Tetragonal | Tetragonal | Tetragonal | Tetragonal |
| Space group | P4/mbm | P4bm | P4bm | P4bm | P4bm | P4bm | P4bm | P4bm |
| RE−O2 (Å) | 2.689(4) | 2.66785(3) | 2.66133(2) | 2.64924(3) | 2.64818(1) | 2.64363(3) | 2.64570(3) | 2.63597(2) |
| RE−O3 (Å) | 2.685(6) | 2.59629(3) | 2.56859(2) | 2.54633(4) | 2.52755(1) | 2.51345(3) | 2.50880(2) | 2.49124(2) |
| RE−RE (Å) | 3.91625(2) | 3.91428(6) | 3.92005(5) | 3.91609(7) | 3.91235(2) | 3.90294(7) | 3.90443(7) | 3.89946(5) |
| O2-RE−O2 (°) | 57.98(12) | 57.441(1) | 57.152(0) | 56.889(1) | 56.930(0) | 56.982(1) | 57.011(1) | 56.831(0) |
| O3-RE−O3 (°) | 180.000(0) | 180.000(0) | 179.992(3) | 180.000(0) | 180.000(0) | 180.000(0) | 180.000(0) | 180.000(0) |
| O3-RE−O3 (°) | 90.000(0) | 90.000(0) | 90.000(0) | 90.000(0) | 90.000(0) | 90.000(0) | 90.00(0) | 90.000(0) |
| O3-RE−O2 (°) | 60.89(17) | 60.439(0) | 60.631(0) | 60.715(0) | 60.619(0) | 60.551(0) | 60.513(0) | 60.623(0) |
| K-O1 (Å) | 2.890(5) | 2.89409(3) | 2.88904(2) | 2.88087(3) | 2.87799(1) | 2.88074(3) | 2.88274(3) | 2.87521(2) |
| K-O2 (Å) | 3.426(5) | 3.45168(3) | 3.45913(2) | 3.46065(4) | 3.45215(1) | 3.45222(3) | 3.45825(3) | 3.45492(2) |
| K-O3 (Å) | 3.516(7) | 3.57919(4) | 3.57673(3) | 3.57304(5) | 3.57185(2) | 3.58052(4) | 3.59013(3) | 3.58091(3) |
| K-O3 (Å) | 3.289(12) | 3.14738(4) | 3.15178(2) | 3.14429(4) | 3.13976(1) | 3.14043(4) | 3.14846(3) | 3.14179(2) |
| K-O4 (Å) | 2.801(7) | 2.79125(3) | 2.79190(2) | 2.78301(3) | 2.78721(1) | 2.78570(3) | 2.78877(3) | 2.78264(2) |
| K-O5 (Å) | 3.061(4) | 3.01536(4) | 3.00598(2) | 2.99839(4) | 2.99106(1) | 2.99650(4) | 3.00095(3) | 2.98983(2) |
| Nb1-O1 (Å) | 1.870(6) | 1.88663(2) | 1.88092(1) | 1.88036(3) | 1.88486(1) | 1.88265(2) | 1.88830(2) | 1.88230(1) |
| Nb1-O2 (Å) | 2.079(8) | 2.05995(3) | 2.06010(1) | 2.04974(3) | 2.03711(1) | 2.03690(3) | 2.03780(2) | 2.03205(2) |
| Nb1-O2 (Å) | 1.984(7) | 1.97891(2) | 1.97080(1) | 1.96450(3) | 1.95911(1) | 1.95722(2) | 1.95988(2) | 1.95070(1) |
| Nb1-O3 (Å) | 1.9677(6) | 1.97711(3) | 1.98281(2) | 1.98223(4) | 1.98143(1) | 1.97894(3) | 1.98081(3) | 1.97940(2) |
| Nb1-O4 (Å) | 1.9598(28) | 1.96594(2) | 1.96655(1) | 1.96530(3) | 1.96014(1) | 1.96277(2) | 1.96769(2) | 1.96380(1) |
| Nb2-O1 (Å) | 1.994(6) | 1.97101(2) | 1.97322(1) | 1.97082(3) | 1.96379(1) | 1.96767(2) | 1.97298(2) | 1.96824(2) |
| Nb2-O5 (Å) | 1.95813(1) | 1.95714(3) | 1.96003(2) | 1.95804(4) | 1.95618(1) | 1.95147(3) | 1.95222(3) | 1.94973(2) |

## ■ RESULTS AND DISCUSSION

### 3.1 Description of crystal Structure.

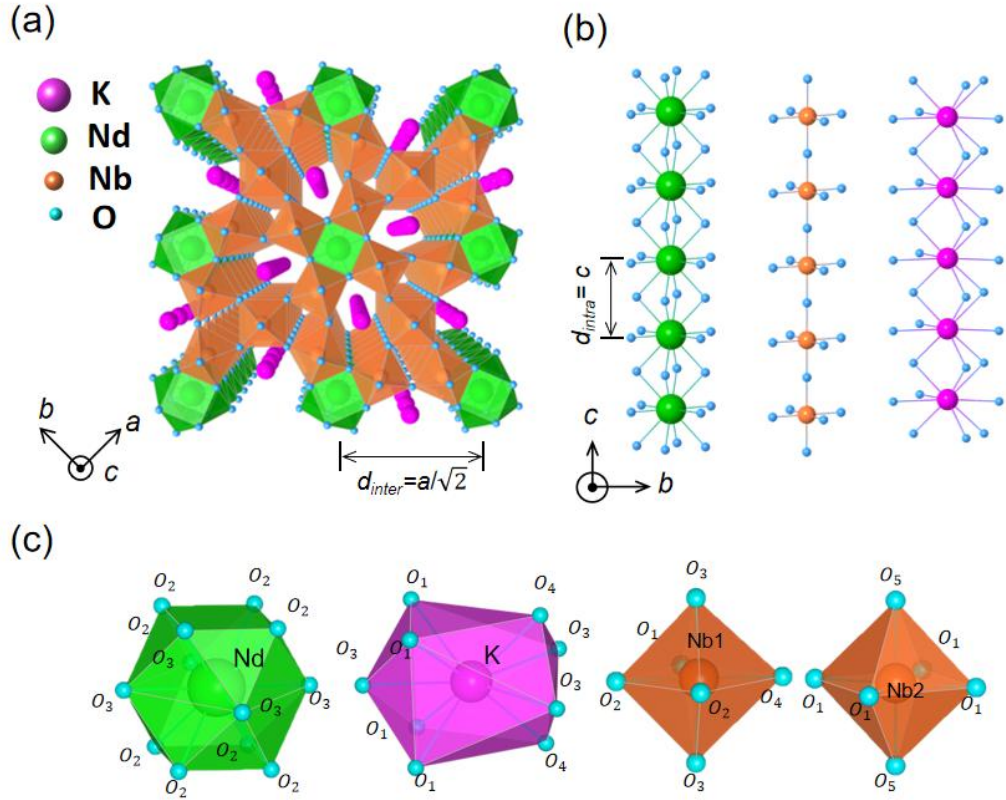

**Figure 1.** (a) The schematic crystal structure of $K_2NdNb_5O_{15}$, the Magenta, green, orange and cyan ball denote K, Nd, Nb and O atoms, respectively. (b) The connections of $NdO_{12}$, $KO_{10}$ and $NbO_8$ polyhedrons along the *c*-axis. (c) The coordination environments of $NdO_{12}$, $KO_{10}$ and $NbO_8$ polyhedrons.

The $K_2RENb_5O_{15}$ (RE=La-Ho) family of compounds are crystallized into the tetragonal tungsten-bronze (TTB) structure,[33,35] most family members undergo a structural phase transition from space group *P*4bm to *I*ma2 then to *P*/4mbm as increased temperature.[38] Considering that the $RE^{3+}$ ions in $K_2RENb_5O_{15}$ have the same lattice geometry independent of the *P*4bm or *P*/4mbm symmetry, as a representative, the room temperature crystal structure of $K_2NdNb_5O_{15}$ is presented as shown in Figure 1a,b. As seen, the structure framework can be regarded to be a network constructed by three columns of linear one-dimensional (1D) cation chains viewed along the *c*-axis, which are connected with the shapes of quadrangular $NdO_{12}$, pentagonal $K_2O_{15}$ and corner-sharing $NbO_6$ octahedra, respectively. The columns of $NbO_6$ octahedra are connected by a corner-sharing fashion with the Nb-O-Nb bond angle of 180° along the *c*-axis, together with all surrounding oxygen atoms they form three different types of channels perpendicular to the *c*-direction. As illustrated in Figure 1b,c, the $K^+$ cations are coordinated by pentacapped pentagonal prims and they build up stands through their trans-oriented basal pentagonal faces along the *c*-axis. In terms of the magnetic lattice topology of $Nd^{3+}$ ions, the 1D uniform chains is formed along the *c*-axis with intrachain separation by $d_{intra}=$ $c$~3.9190(2) Å and well spatially separated by the NbO$_6$ octahedron with interchain distance $d_{inter}=a/\sqrt{2}$ ~ 8.8424(2) Å, here the interchain distance is larger than the ones of 1D spin chain Nd(HCOO)$_3$ and Ca$_4$NdO(BO$_3$)$_3$ compounds.[39,22] Additionally, here the columns of Nd$^{3+}$ spin chains are located on a perfect square lattice within the *ab*-plane, as shown in Figure 1c and Figure S1 in supporting information, this is different from the frustrated triangular lattice arrangement in RE(HCOO)$_3$ and Ca$_4$REO(BO$_3$)$_3$ systems.[20,23] From this viewpoint, K$_2$NdNb$_5$O$_{15}$ forms a linear 1D spin chain structure with low dimensionality but without geometric frustration.

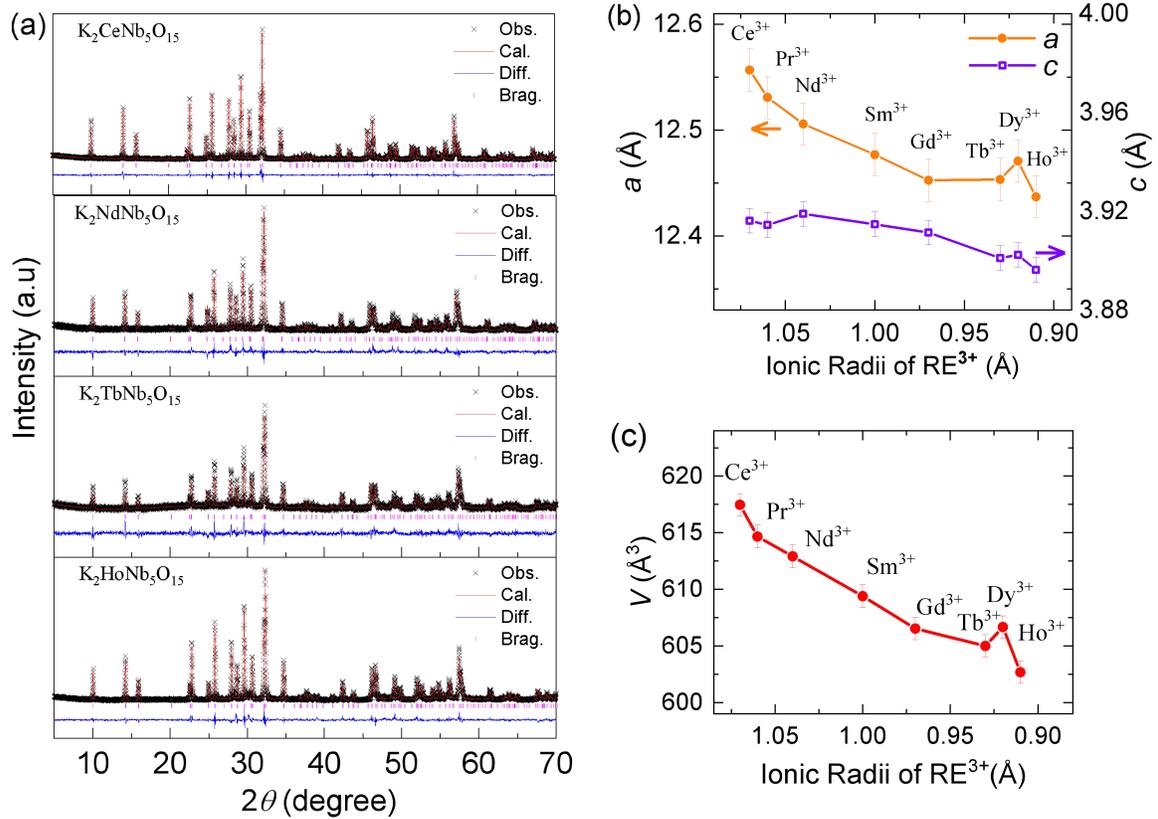

**Figure 2.** (a) Room-temperature powder X-ray diffraction (XRD) spectra of K$_2$RENb$_5$O$_{15}$(RE=Ce, Nd, Gd, Ho) samples: the black cross denote the experimental data, red and blue lines are the calculated patterns and its difference with the experimental data, the pink stick reflect the Bragg reflection positions. (b) The lattice parameters (*a,c*) versus the ionic radii of RE$^{3+}$ ions in K$_2$RENb$_5$O$_{15}$. (c) The variation of unit-cell volume (V) as a function of ionic radii of RE$^{3+}$ ions in K$_2$RENb$_5$O$_{15}$.

In the unit cell of K$_2$NdNb$_5$O$_{15}$, there exist nine crystallographic sites including one unique Nd atom (Wyckoff site, 2a), one K atom (Wyckoff site, 4g), two Nb atoms (Wyckoff site, 2c, 8j) and five

O atoms. The detailed coordination of $Nd^{3+}$, $K^+$, $Nb1^{5+}$ and $Nb2^{5+}$ cations with surrounding oxygen ions are depicted on Figure 1c. Based on the structural refinements of powder x-ray diffraction (XRD) data, the obtained interatomic distances and bond angles are listed in Table 1. In the pentagonal $K_2O_{15}$ polyhedron, the K-O bond lengths are ranging from 2.788(11)-3.419(16) Å. For the $[NbO_6]^{7-}$ octahedra, the Nb1 is coordinated with large distortion by six oxide anions with distances of 1.870–2.079 Å whereas the octahedron about Nb2 has a less distorted environment with typical distances of 1.9572 Å and 1.976 Å (see Table 1 for interatomic distances). The magnetic $Nd^{3+}$ ions are coordinated into a shape of tetracapped-cube-like geometry with $D_{4h}$ point group symmetry, where four capping oxide anions $(O2)^{2-}$ are located above and below the rectangular faces of the square-prismatic $(O3)^{2-}$ polyhedron. Four $Nd-O_3$ bond lengths have same distance of 2.750(14) Å and eight $Nd-O_3$ bond length are also equidistant with 2.774(10) Å. For the magnetic exchange interactions between $Nd^{3+}$ ions, the intrachain super-exchange pathway are realized via a tetrameric structure connections with the neighbouring $NdO_{12}$ tetrakaidecahedron along the c-axis, while the nearest interchain ones are through the Nd-O-K-O-Nd pathway without co-sharing oxygen ions within the ab-plane. Here, the $Nd^{3+}$ spin chains are well separated by the nonmagnetic K/Nb-O polyhedrons with large distances as ideal nonmagnetic spacers. Since the dipolar energies scale as $1/r^3$,[40] the intrachain nearest-neighbor (NN) dipole–dipole interactions are ~11.5 times larger than the NN interchain interactions, this supports $K_2NdNb_5O_{15}$ to be a quasi-1D spin chain system.

Room temperature experimental and refined powder XRD patterns for representative $K_2RENb_5O_{15}$ (RE=Ce,Nd,Tb,Ho) samples are shown in Figure 2a. For RE=Ce, structure refinement is based on the aristotype cell of TTB structure with space group P/4mbm (no. 127), while for RE=Nd,Tb,Ho, the space group P4bm (no. 100) is used. The crystal structure adopted for structure refinement is supported by the differential scanning calorimetry (DSC) [see Figure S2] and temperature-dependent dielectric measurement results. As shown in Figures. 3a-c, temperature dependence of dielectric constant $\varepsilon_r(T)$ at selected frequencies were measured on three typical $K_2RENb_5O_{15}$ (RE= Ce, Nd, Gd) samples. For RE= Nd, Gd, the observed two anomalies at $T_{C1}$ and $T_{C2}$ in the $\varepsilon_r(T)$ curves are related to the phase transition from ferroelectric (FE) to antiferroelectric(AFE) to paraelectric (PE) phase, respectively. Also, these two phase transitions are also revealed by the DSC curves [see Figure. S2] in consistent with the previous report.[34] Thus, room temperature crystal structure of $K_2RENb_5O_{15}$ (RE= Nd, Gd) belongs to the ferroelectric P4bm space group. For $K_2CeNb_5O_{15}$, as shown in Figure 3a, two phase transitions from FE to AFE to PE phase occur at $T_{c1}$=130 K and $T_{c2}$=250 K. Then, $K_2CeNb_5O_{15}$ crystallizes into the crystal structure with P4bm symmetry and P/4mbm at temperatures $T < T_{c1}$ and $T > T_{c2}$, respectively.

The calculated XRD patterns on the serial $K_2RENb_5O_{15}$ compounds match well with the experimental data, and the refined lattice parameters are summarized in Table S1. As expected, the lattice parameters and unit-cell volume monotonically decrease with the reduced $RE^{3+}$ ionic radii as shown in Figures 2b,c. For the detailed structure parameters, the bond distances, bond angles, the intraplane and interplane RE−RE distances of the four compounds are summarized in Table 1. From the structure analysis, no distinct antisite mixing occupancy between the magnetic RE$^{3+}$ ions and nonmagnetic K$^+$/Nb$^{5+}$ cations is detected, which can be due to the differences of their ionic radii and coordinate environments. Here, the disorder-free character in K$_2$RE$_2$Nb$_5$O$_{15}$ is important to investigate the intrinsic ordered 1D spin chain physics, which avoid the puzzle on clarifying its magnetic ground state driven by magnetic exchange randomness as the report in quasi-1D Ca$_4$REO(BO$_3$)$_3$[23] as well as the quantum spin liquid YbMgGaO$_4$ candidate.[41,42]

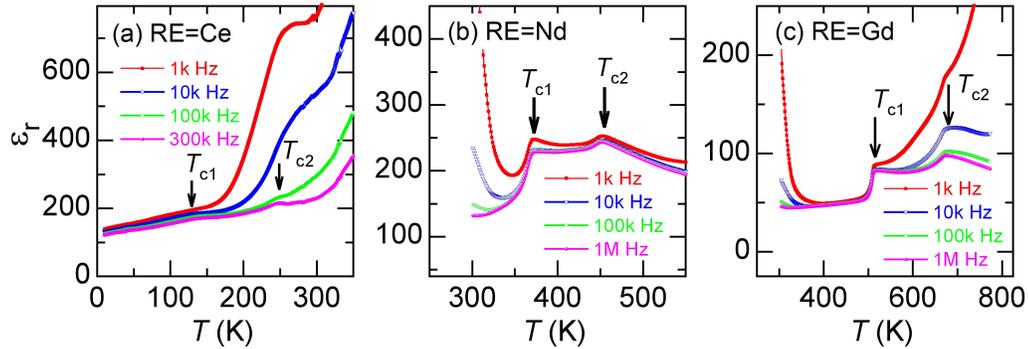

**Figure 3.** Temperature dependence of dielectric constant for K$_2$RENb$_5$O$_{15}$ with (a) RE=Ce (b) RE=Nd and (c) RE=Gd, respectively.

### 3.2 Magnetic properties and electron spin resonance results

Temperature ($T$) dependence of magnetic susceptibilities $\chi(T)$ for all serial K$_2$RENb$_5$O$_{15}$ (RE=Ce-Ho) compounds are measured from 1.8 K to 300 K in an applied field $\mu_0H$ =0.1 T, as presented in Figure 4. The inverse susceptibilities 1/$\chi(T)$ are fitted by the Curie-Weiss (CW) law $\chi = C/(T-\theta_{\mathrm{CW}})$, where $\chi$ is the susceptibility, $C$ is the Curie constant and $\theta_{\mathrm{CW}}$ is the Curie-Weiss temperature. The effective magnetic moments $\mu_{\mathrm{eff}}$ are obtained by the following relationship: $\mu_{\mathrm{eff}} = (3k_{\mathrm{B}}C/N_{\mathrm{A}})^{1/2}$, where $k_{\mathrm{B}}$ is the Boltzmann constant and $N_{\mathrm{A}}$ is the Avogadro's number. Considering that the crystal electric field (CEF) splitting can affect the $\chi(T)$ behaviors due to the variable occupied population of electrons at different CEF multiplets,[43,44] the CW fits are performed at both high and low temperature regimes, the resultant $\theta_{\mathrm{CW}}$ and $\mu_{\mathrm{eff}}$ are summarized in Table 2, where the effective moments ($\mu_{\mathrm{fi}}$) of free RE$^{3+}$ ions calculated by $g_J[J(J+1)]^{1/2}$ are also presented for comparison. The isothermal field-dependent magnetization $M$ ($\mu_0H$) curves were measured at selected temperatures ($T$ < 20 K), and high-field magnetization characterizations ($\mu_0H$ ~50 T) on K$_2$RENb$_5$O$_{15}$ (RE=Pr, Sm, Ho) were performed to determine the saturated magnetization [ see Figure 6]. Additionally, the X-band electron spin resonance (ESR) spectra were measured at temperatures below 30 K, the derivative ESR spectra (d$P$/d$\mu_0H$) ($P$ is the integral ESR intensity) for K$_2$RENb$_5$O$_{15}$ are presented in

Figure 7. From that, the g-factors are calculated as $g = h\nu/\mu_B\mu_0 H_r$, where $h$ is the Planck constant, $\nu$ = 9.4 GHz is the microwave frequency, $\mu_0 H_r$ is the resonance absorption field and $\mu_B$ is the Bohr magneton. The obtained temperature-dependent Lande g-factors for $K_2RENb_5O_{15}$ (RE=Ce, Nd, Gd, Dy) compounds are shown in Figure 8. The elementary magnetic properties for $K_2RENb_5O_{15}$ are descried separately below.

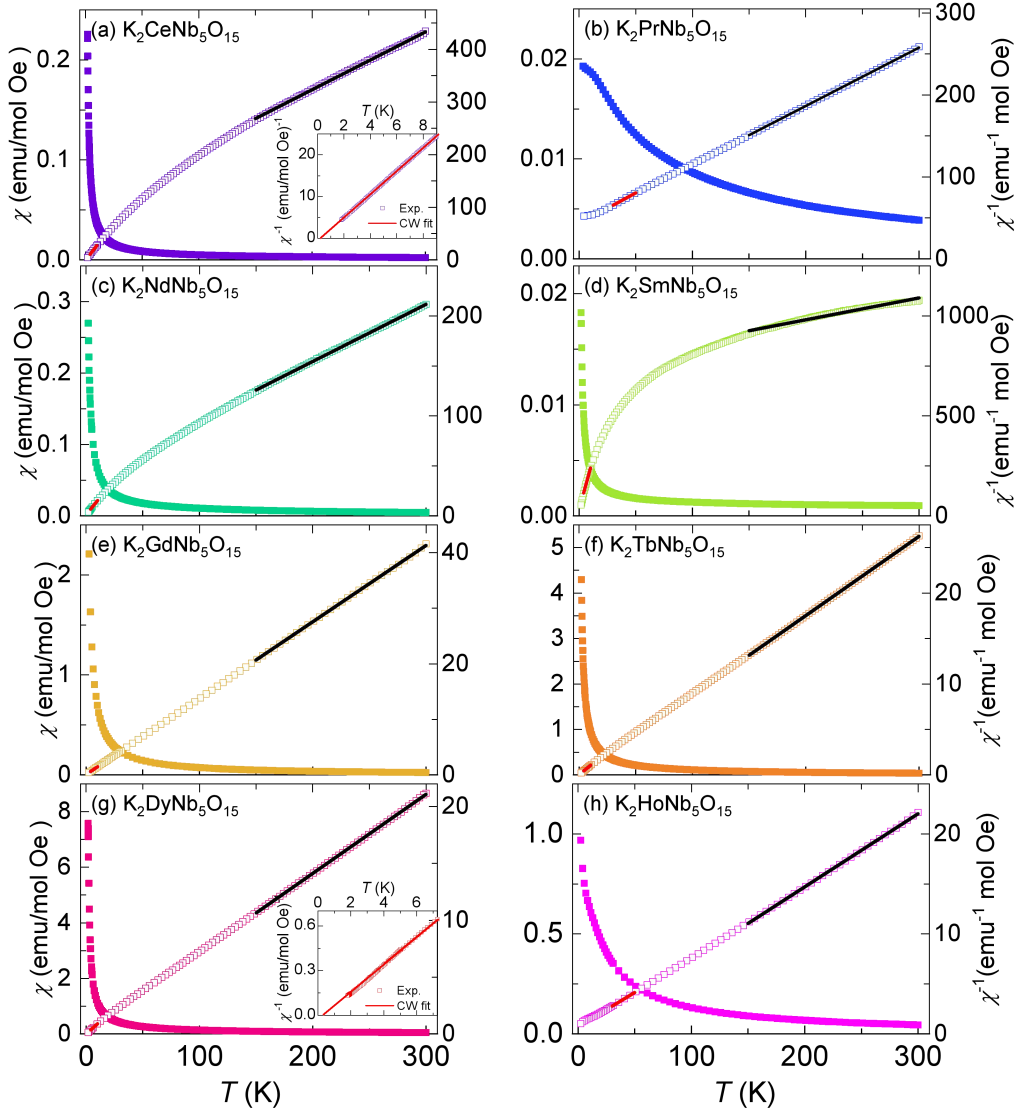

**Figure 4.** Temperature dependence of magnetic susceptibility $\chi(T)$ and inverse magnetic susceptibility $\chi^{-1}(T)$ under field of $B$=0.1 T for $K_2RENb_5O_{15}$ (RE=Ce, Pr, Nd, Sm, Gd-Ho) samples. The solid black and red lines show the CW fits at high and low temperature regimes, the inset in (a) and (g) show the enlarged region of $\chi^{-1}(T)$ at low temperatures.

**Table 2.** The Curie-Weiss temperatures ($\theta_{CW}$) and effective magnetic moments ($\mu_{eff}$) determined from the Curie-Weiss fitting at low and high temperatures to magnetic susceptibility $\chi(T)$ of $K_2RENb_5O_{15}$(RE =Ce—Ho) compounds, the effective moment ($\mu_{fi}$) for free ions are calculatedy$[J(J + 1)]^{1/2}$

| RE | High $T$ fit | $\theta_{CW}$(K) | $\mu_{eff}(\mu_B)$ | Low $T$ fit | $\theta_{CW}$(K) | $\mu_{eff}(\mu_B)$ | $\mu_{fi}(\mu_B)$ |
|---|---|---|---|---|---|---|---|
| Ce | 150 -300 K | -94.5 | 2.7 | 4-10 K | 0.183 | 1.70 | 2.54 |
| Pr | 150 -300 K | -62.2 | 3.37 | 30-50 K | -57.3 | 3.26 | 3.58 |
| Nd | 150 -300 K | -70.3 | 3.74 | 4-10 K | -0.87 | 2.38 | 3.62 |
| Sm | 150 -300 K | -400.0 | 2.70 | 4-10 K | -1.56 | 0.63 | 0.84 |
| Gd | 150 -300 K | -0.52 | 7.64 | 4-10 K | -0.61 | 7.62 | 7.94 |
| Tb | 150 -300 K | - 0.92 | 9.57 | 4-10 K | - 0.39 | 8.94 | 9.72 |
| Dy | 150 -300 K | -2.91 | 10.71 | 4-10 K | 0.185 | 9.36 | 10.63 |
| Ho | 150 -300 K | -1.23 | 10.45 | 30-50 K | -10.5 | 10.8 | 10.60 |

**$K_2CeNb_5O_{15}$:** Temperature-dependent susceptibility $\chi(T)$ of $K_2CeNb_5O_{15}$ is shown in Figure 4a, no long-range magnetic order or spin freezing is detected down to 1.8 K. High temperature CW fits to $\chi^{-1}(T)$ curves give rise to $\theta_{CW}$ = -95.3 K and $\mu_{eff}$ = 2.70$\mu_B$/Ce$^{3+}$. The effective moment is close to the value $g_J[J(J+1)]^{1/2}$ =2.54 $\mu_B$ for free Ce$^{3+}$ ($^2F_{5/2}$) ions. At low temperatures, the CW fits yield $\theta_{CW}$ =0.181 K and effective moment $\mu_{eff}$ = 1.55 $\mu_B$ /Ce$^{3+}$. Here, the positive value of $\theta_{CW}$ indicates dominant ferromagnetic (FM)-like interaction between the Ce$^{3+}$ local moments. As shown in Figure 5a, the isothermal magnetization $M$ ($\mu_0H$) curves at 2 K display a nonlinear field dependence behaviors up to field of 7 T. The saturated magnetization ($M_S$) reaches ~0.78 $\mu_B$ /Ce$^{3+}$ is much smaller than the expected value $M_S$ = $g_JJ\mu_B$ = 2.1$\mu_B$ for free Ce$^{3+}$ ions but close to the half value of low-$T$ fitted effective moment of $\mu_{eff}$ = 1.70 $\mu_B$/Ce$^{3+}$.

The measured ESR spectra at selected temperatures for $K_2CeNb_5O_{15}$ are shown in Figure 7a. From that, we can identify two broad resonance lines with peaks at $\mu_0H_{r1}$ and $\mu_0H_{r2}$, respectively. The low-field broad central line splits into several lines, possibly due to the non-equivalent paramagnetic centers as reported in the ESR spectra of $NdMgAl_{11}O_{19}$ and $Nd_3Ga_5SiO_{14}$ compounds with low point group symmetry. [45,46] Based on the position of resonance lines, two effective g-factors ($g_1$~3.67 and $g_2$~1.26) can be obtained, and their temperature dependence behaviors are shown in Figure 8a. At 2 K, the calculated powder averaged g-factor is $g_{ave} = \frac{1}{3}g_1 + \frac{2}{3}g_2$~2.05. In case of $j_{eff}$=1/2 for Ce$^{3+}$ moment, the low-$T$ calculated $\mu_{eff} = g_{ave}\sqrt{j_{eff}(j_{eff} + 1)}$=1.78 $\mu_B$/Ce$^{3+}$ is close to the low-$T$ fitted $\mu_{eff}$ = 1.70 $\mu_B$/Ce$^{3+}$.

**$K_2PrNb_5O_{15}$:** Figure 4b presents the magnetic susceptibility results of $K_2PrNb_5O_{15}$, high temperature CW fitting on $\chi^{-1}(T)$ curves yields $\mu_{eff}$ = 3.37$\mu_B$/Pr$^{3+}$ and this is close to the value of 3.58$\mu_B$ for free Pr$^{3+}$ ($4f^2,^3H_4$) ion. As $T$ is below 30 K, 1/$\chi(T)$ starts to deviate from the linear temperature dependence and exhibits a broad hump-shaped behaviour. The isothermal $M$ ($\mu_0H$) curve at 2 K exhibits a relative small magnetization ~0.23 $\mu_B$ at 7 T [see Figure 5b], which can be due to the formation of well-separated singlet ground state by the CEF splitting as report in other non-Kramers $Pr^{3+}$ ($4f^2$, $J= 4$) systems.[44,47] Also, the very weak anomalous peaks detected from the ESR curves suggest a nonmagnetic singlet ground state [see Figure S3]. Moreover, high field magnetization at 2 K shows a linear field dependence behaviour without saturation up to $\mu_0H$ = 50 T [ see Figure 6a], the maximum magnetization ~1.51$\mu_B$/Pr is still much smaller than the saturated magnetization $M_S = g_JJ\mu_B = 3.2\mu_B/Pr^{3+}$ for free $Pr^{3+}$ ions.

**$K_2NdNb_5O_{15}$:** The magnetic susceptibility of $K_2NdNb_5O_{15}$ is shown in Figure 4c, high-temperature CW fits on $\chi^{-1}(T)$ curves result in $\theta_{CW}$ = -70.3 K and $\mu_{eff}$ = 3.74 $\mu_B/Nd^{3+}$. The obtained moment is close to the value of free $Nd^{+3}$ ($4f^3$, $^4I_{9/2}$) ions (3.62 $\mu_B$). At low temperatures, the reduced effective moment $\mu_{eff}$ = 2.38 $\mu_B/Nd^{+3}$ is obtained due to the CEF effect, where electrons are apt to occupy the low-lying doublets states at low temperatures. As shown in Figure 5c, the isothermal $M(\mu_0H)$ at 2 K exhibits the saturated magnetization $M_s$~1.38 $\mu_B/Nd^{+3}$ at field of 7 T, this value is slightly large than the half of effective magnetic moment. As shown in Figure 7b, single ESR resonance peak is observed and from that the $g$-factor $g$=2.5 at 2 K is obtained [see Figure 8b]. Then, the saturated magnetization of powders can be estimated to $M_{sat} = gJ_{eff}\mu_B \sim 1.25\mu_B$, this is slightly smaller than the experimental $M(\mu_0H)$ results.

**$K_2SmNb_5O_{15}$:** The magnetic susceptibility of $K_2Sm_2Nb_5O_{15}$ is shown in Figure 4d. As seen, the $\chi(T)$ exhibits a characteristics of Van-Vleck type paramagnetic (PM) contribution as usually observed in the Sm-contained oxides.[48,49] The low temperature (4- 20 K) Curie–Weiss fitting gives $\theta_{CW}$ = -1.56 K and $\mu_{eff}$=0.63 $\mu_B/Sm^{3+}$. As shown in Figure 5d, the isothermal $M$ ($\mu_0H$) curves at 2 K nearly show linear field dependences up to 7 T without saturation. Based on the high field magnetization measurement as shown in Figure 6b, the maximum magnetization value 0.59 $\mu_B$ at $\mu_0H$ ~ 50 T is close to the low-$T$ fitted magnetic moment in case of formation of Kramers doublet ground state of $Sm^{3+}$ ions.

**$K_2GdNb_5O_{15}$:** The magnetic susceptibility curve of $K_2GdNb_5O_{15}$ is shown in Figure 4e, high-temperature susceptibility CW fits give $\mu_{eff}$ = 7.64 $\mu_B/Gd^{3+}$ and $\theta_{CW}$ = −0.54 K, and the low temperature fits (8 $K$ <T< 30 $K$) yield $\mu_{eff}$ = 7.62 $\mu_B/Gd^{3+}$ and $\theta_{CW}$ = −0.62 K. The obtained moments agree with the free-ion value of 7.94 $\mu_B$ for $Gd^{3+}$ ($4f^7$,$^8S_{7/2}$) ions, and here negative $\theta_{CW}$ reveal the dominant AFM exchange interactions between $Gd^{3+}$ moments. As shown in Figure 5e, the saturated magnetization $M_S$ ~ 6.9 $\mu_B$ at 2 K is close to the value $gJ\mu_B$=7.0 $\mu_B$ of $Gd^{3+}$ ($4f^7$,$S$=7/2, $L$=0) ions. Moreover, single resonance lines can be identified from the resonance peaks of ESR spectra at different temperatures as shown in Figure 7c. The estimated $g$-factor have the values of 2.32 to 2.41 in temperature range from 30 K to 2 K [see Figure 8c], which is slightly larger than the value of free $Gd^{3+}$ ions.

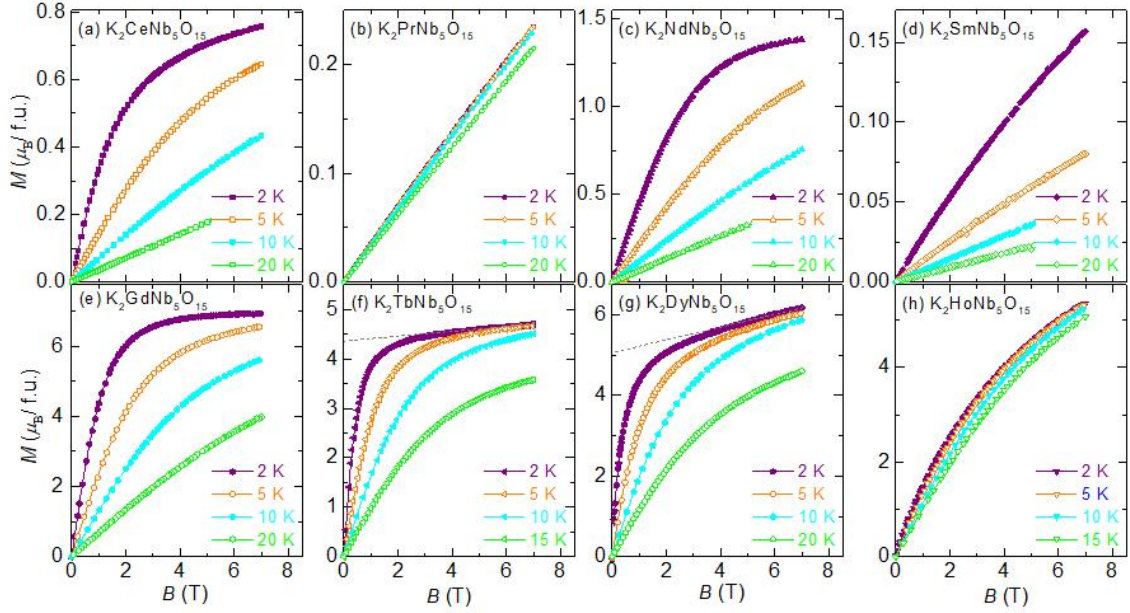

**Figure 5.** Field dependence of magnetization $M$ ($B$) curves as selected temperatures for $K_2RENb_5O_{15}$ (RE=Ce, Pr, Nd, Sm, Gd-Ho).

**$K_2TbNb_5O_{15}$:** The magnetic susceptibility curve of $K_2Tb_2Nb_5O_{15}$ is shown in Figure 4f, no magnetic ordering is detected with temperature down to 1.8 K. High temperature CW fitting gives rise to $\theta_{CW}$ = −0.92 K and $\mu_{eff}$ = 9.57 $\mu_B/Tb^{3+}$, this moment value is slightly smaller than the moment 9.72 $\mu_B$ for free $Tb^{+3}$ ions ($4f^8$, $^7F_6$) with $m_J= \pm 6$ doublet and $g$ = 3/2. At low temperatures (4 K< $T$ < 10 K), the CW analysis yields $\theta_{CW}$ = −0.39 K and $\mu_{eff}$ = 8.94 $\mu_B/Tb^{3+}$, here the negative $\theta_{cw}$ suggests the dominant AFM interactions. As shown in Figure 5f, the isothermal $M(B)$ curves at 2 K exhibit nonlinear field dependence behavior at low fields ($\mu_0H$ <1.5 T) and then saturates above 2 T. The saturated magnetization $M_S$~4.7 $\mu_B$ is close to half of expected saturated moment 10.31 $\mu_B$. As shown in Figure S3, single resonance lines can be found at low field located at ~350 Oe, which reveals a large $g$-factor value.

**$K_2DyNb_5O_{15}$:** As shown in Figure 4g, high temperature CW fitting on $\chi^{-1}(T)$ of $K_2Dy_2Nb_5O_{15}$ gives $\theta_{CW}$ = -2.91 K and $\mu_{eff}$ = 10.71 $\mu_B/Dy^{3+}$, this value is in agreement with 10.63 $\mu_B/Dy^{3+}$ of free $Dy^{3+}$($4f^9$,$^6H_{15/2}$) ion. As comparison, similar fits at low temperatures (4 -10 K) lead to positive $\theta_{CW}$= 0.185 K and reduced moment $\mu_{eff}$ = 9.36 $\mu_B/Dy^{3+}$, as shown in the inset of Figure 4g. The $M(H)$ curves at 2 K show a nonlinear behaviour at low fields ($B$ < 3 T) and above that it has linear field dependence possibly originated from the Van-Vleck PM contribution as shown in Figure 5g. After subtracting the Van Vleck PM component, the saturated magnetization ~ 5.1 $\mu_B/Dy^{3+}$ reaches the half of full polarized $Dy^{3+}$ moments with $g_J J\mu_B$ = 10 $\mu_B$ of free $Dy^{3+}$ ions, similar magnetization value is also observed in pyrochlore-lattice $Dy_2Ti_2O_7$ and triangular lattice $Dy(BaBO_3)_3$.[50,51] As shown in Figure 7d, two resonance ESR lines can be found at different temperatures and the Lande g-factor can be estimated from the resonance peaks, the obtained values are shown in Figure 8d. Using the obtained g-factor $g_1$=5.1 and $J_{eff}$=1/2 at 2 K, the estimated magnetization $M_{sat} = gJ_{eff}\mu_B \sim 2.5\mu_B$ is much smaller than the experimental value, which suggests the existence of magnetic contribution from the orbital moments of $Dy^{3+}$ ions.

**$K_2HoNb_5O_{15}$:** The magnetic susceptibility of $K_2HoNb_5O_{15}$ is shown in Figure 4h, no long-range magnetic ordering is detected down to 2 K. The high-temperature CW fits to $\chi^{-1}(T)$ give rise to $\theta_{CW}$ = -1.23 K and $\mu_{eff}$ = 10.45$\mu_B$/$Ho^{3+}$. The effective moment is close to the value $g_J[J(J + 1)]^{1/2}$ =10.6$\mu_B$ for free $Ho^{3+}$ ($4f^{10}$,$^5I_8$) ions with $m_J$ = ±8 doublet ground state. As decreased temperature $T$ < 30 K, the $1/\chi(T)$ deviates from the linear temperature dependences with an upturn behaviour. As shown in Figure 5b, the isothermal $M(B)$ curve at 2 K don't reach saturation up to $B$ = 7 T. As field increases up to 50 T, magnetization reaches maximum with value of 10.2$\mu_B$ [see Figure 6c], which is close to the low-temperature fitted moment.

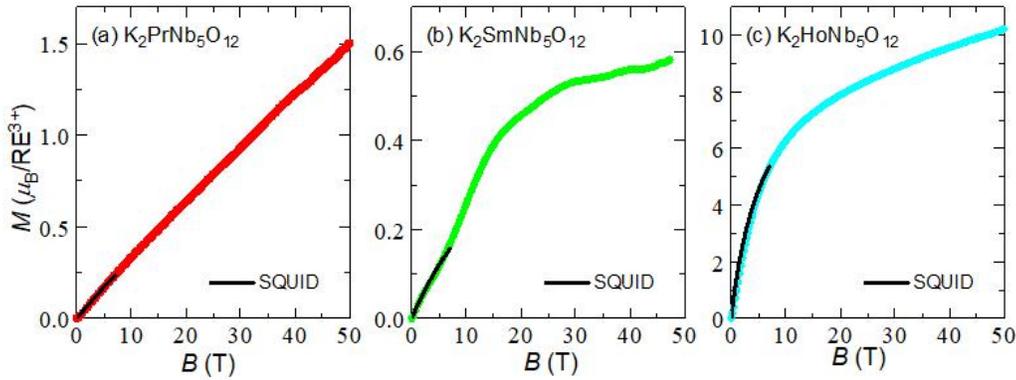

**Figure 6.** High field magnetization data of $K_2RENb_5O_{15}$ (RE= Pr, Sm, Ho) at 2 K with $B$ up to 50 T.

The series of $K_2RENb_5O_{15}$ compounds can be divided into two categories: containing the Kramers RE ions with odd number of 4f electrons ($Ce^{3+}$,$Nd^{3+}$,$Sm^{3+}$,$Gd^{3+}$ and $Dy^{3+}$) and non-Kramers ions with even number of 4f electrons ($Pr^{3+}$,$Tb^{3+}$ and $Ho^{3+}$). Among them, $Gd^{3+}$ is special and has half-filled 4f shell ($4f^7$, $S$=7/2, $L$= 0) with spin-only effective magnetic moment $S_{eff}$=7/2, then the negative value of $\theta_{CW}$ = −0.62 K suggests the AFM interactions in $K_2GdNb_5O_{15}$. For the members with Kramers ions, the low-temperature magnetism can be described by $j_{eff}$ =1/2 state protected by time-reversal symmetry. While for compounds with non-Kramers ions, only in case of the formation of two low-lying CEF singlets separated by a small energy gap, their magnetic behaviors can be described by pseudospin $S_{eff}$ = 1/2 state at low temperatures, but here the local $C_{4h}$ symmetry environments in $REO_{12}$ polyhedra result in the formation of well separated two lowest-lying CEF singlets as indicated by the small magnetization at field of 7 T and 2 K. To determinate the CEF level splitting, the inelastic neutron scattering experiment is required and which can help to clarify the single-ion anisotropy and low temperature magnetic ground states.

**Table 3.** The estimated intra-chain (Dintra-chain) and inter-chain (Dinter-chain) dipole interactions and nearest-neighbor exchange ($J_{nn}$) interactions in $K_2RENb_5O_{15}$ (RE=Ce, Pr, Nd, Sm, Gd, Tb, Dy and Ho) compounds.

| RE | $J_{nn}$ using $J$ (K) | $J_{nn}$ using $S_{eff} = 1/2$ (K) | $D_{inter-chain}$ (K) | $D_{intra-chain}$ (K) |
|---|---|---|---|---|
| Ce | 0.0157 | 0.183 | -0.0051 | -0.06 |
| Pr | -2.15 | N/A | -0.018 | -0.22 |
| Nd | -0.0264 | -0.87 | -0.0102 | -0.117 |
| Sm | -0.134 | -1.56 | -0.00072 | -0.0082 |
| Gd | -0.042 | N/A | 0.106 | -1.21 |
| Tb | -0.00696 | 0.39 | -0.146 | -1.68 |
| Dy | 0.00218 | 0.185 | -0.159 | -1.838 |
| Ho | 0.109 | N/A | -0.213 | -2.456 |

To well describe the 1D characteristics of $K_2RENb_5O_{15}$ compounds, we evaluate the magnitude of superexchange magnetic interactions and the intrachain and interchain dipolar interactions. The strength of superexchange interactions between local $RE^{3+}$ moments are estimated by the mean field approximation using $J_{nn}$ = $3k_B\theta_{CW}$ /$zS_{eff}(S_{eff}+1)$,[52] where $S_{eff}$ denotes the total spin quantum number, and z is the number of nearest-neighbouring spins (here n = 2). Considering the presence of large SOC, diverse CEF effect and spin types of $RE^{3+}$, both the quantum number $J= |L \pm S|$ of $RE^{3+}$ ions and effective spin number $S_{eff}$=1/2 are employed to estimate the nearest-neighbour exchange $J_{nn}$, the results are shown in Table 3. For compounds with Kramers doublet state, it is more suitable to take $S_{eff} = 1/2$ and that is applicable to the $K_2RENb_5O_{15}$ (RE=Ce, Nd, Sm, Dy), where the electrons will occupy on the low-lying Kramers doublet ground state at low temperatures due to its well separation from the first CEF excited level. For the compounds with quasi-doublet ground state well separated from other excited CEF levels, like the case in TmMgGaO$_4$ and TbInO$_3$ compounds where $S_{eff} = 1/2$ is suitable.[53,54] The dipolar interaction (D) is evaluated by D=

$\mu_0\mu_{eff}^2/4\pi(r_{nn})^3$,[40] where $r_{nn}$ can be the nearest neighboring (NN) intra- and inter-chain RE-RE distances ($d_{intra}$, $d_{inter}$), $\mu_{eff}$ is the low-$T$ fitted effective moment. The calculated dipolar interactions are also listed in Table 3. From that, the intra-chain dipolar interactions ($D_{intra}$) is ~10 times larger than the inter-chain ones ($D_{inter}$). As one example, for $K_2NdNb_5O_{15}$, the inter-chain dipole interaction ($D_{intra}$~-0.117 K) is more than ~10 times larger of the interchain dipole interaction ($D_{inter}$~-0.0102 K). From the structure viewpoint, due to the insulating nature of $K_2RENb_5O_{15}$, the interchain super-exchange magnetic interactions through the RE-O-K/Nb-O-RE pathways should be much smaller than that of intrachains' ones via the RE-O-RE pathways. All the estimated interactions indicate that $K_2RENb_5O_{15}$ realize a quasi 1D magnetism.

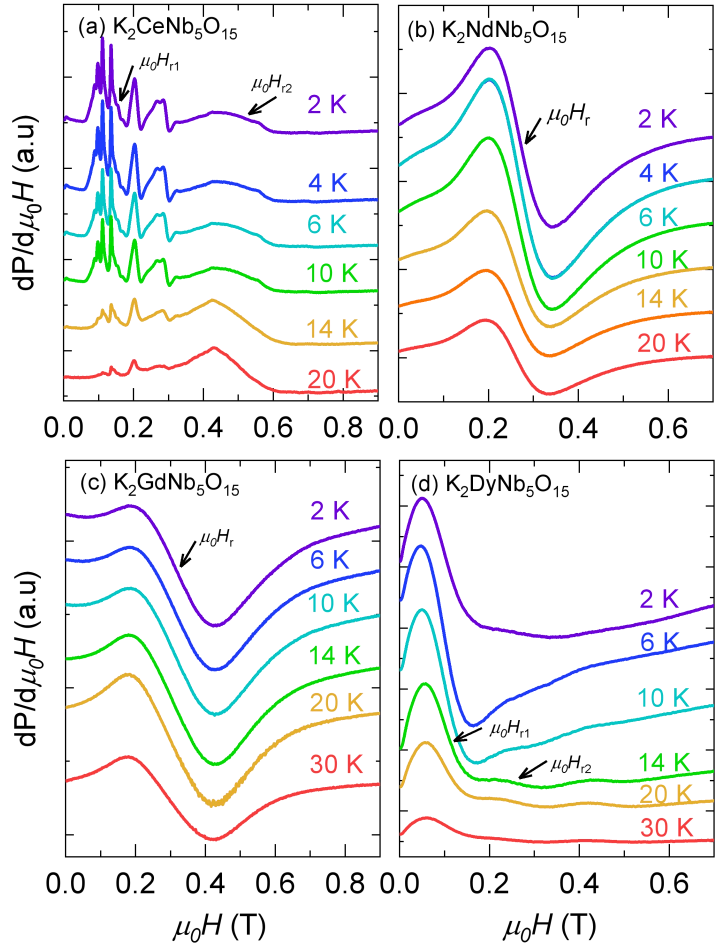

**Figure 7.** The X-band electron spin resonance (ESR) spectra at selected temperatures for $K_2RENb_5O_{15}$ (RE= Ce, Nd, Gd, Dy).

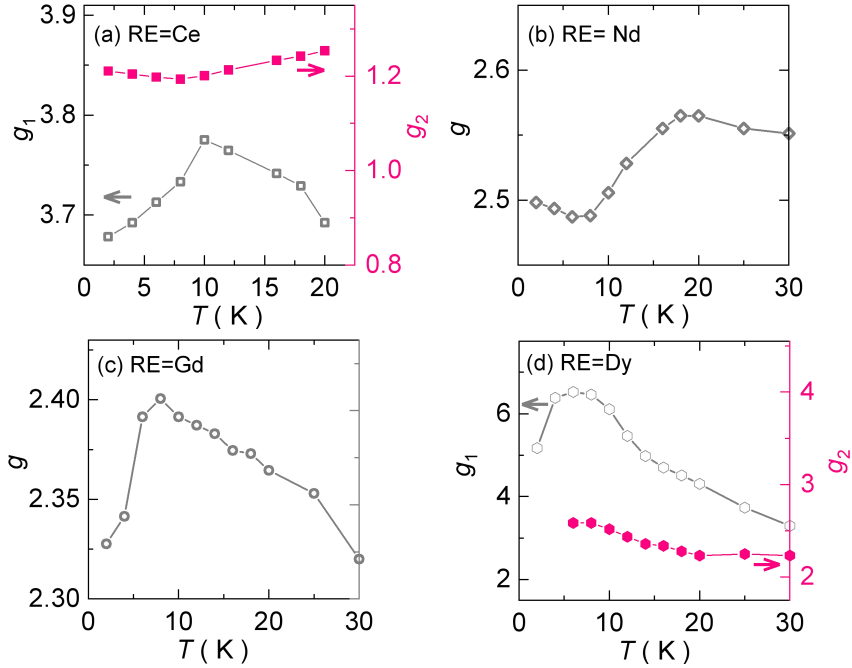

**Figure 8.** Temperature dependence of $g$-factors for $K_2RENb_5O_{15}$ (RE= Ce, Nd, Gd, Dy).

### 3.3 Low-temperature specific heat and magnetocaloric effect of $K_2GdNb_5O_{15}$

The low dimensional RE-based magnets are good candidate materials to realize sub-Kelvin cooling by the adiabatic demagnetization (ADR) procedure, this gain strong interest for its important applications in supplying sub-Kelvin environment for space technology and quantum computers.[27,28] Compared to the compounds with $S_{eff}$=1/2 moment, the isostructural Gd-based ones ($S_{eff}$=7/2) can possess three times of entropy storage capacity and long holding time. In this case, below we focus on the low-$T$ magnetocaloric (MC) effect of $K_2Gd_2Nb_5O_{15}$. Indeed, compared to the other $K_2RE_2Nb_5O_{15}$ members with relatively large magnetizations for RE=Tb,Dy,Ho, $K_2Gd_2Nb_5O_{15}$ exhibit the optical MC effect in temperature region from 2 -12 K, as shown in Figure 9a. The magnetic entropy changes ($\Delta S_m$) in field changes ($\Delta B$=0T -7 T) were calculated from the isothermal magnetization curves by employing Maxwell's relation:[55] $\Delta S_m = \int_{B_i}^{B_f} \left(\partial M/\partial T\right)_B dB$, where $B_i$ (usually $B_i$=0) and $B_f$ represent the initial and final values of magnetic field, respectively. As decreased temperatures, temperature-dependent $\Delta S_m$ at different fields show the monotonic increase [see Figure 9b], which indicate the maximum MC effect appeared at $T$ < 2 K.

To unveil the magnetic ground state and MC effect of $K_2GdNb_5O_{15}$ at low temperatures, we performed the specific heat $C_p(T)$ measurements at different fields down to 0.08 K. As shown in Figure 10a, zero-field $C_p(T)$ curves show a sharp peak at $T_N$~0.31 K signifying the formation of long-range magnetic order. Also, a broad shoulder with maximum at $T_{sr}$~0.45 K is detected, this is distinct from a single λ-like peak usually observed in classical magnets, such as $GdAlO_3$ and $Gd_2Sn_2O_7$ compounds.[56,57] After subtracting the lattice contribution ($C_{Latt}$) using the nonmagnetic analogue $K_2La_2Nb_5O_{15}$ as referee, magnetic specific heat $C_M(T)$ curves are plotted in Figure 10b. Further integrating $C_{mag}/T$ over temperature, the obtained magnetic entropy $S_M(T)$ is also presented in Figure 10a. As increased $T$, $S_M(T)$ approaches ~94% of $R\ln(2S+1)=8R\ln8$ up to 20 K as expected for the $S_{eff}$=7/2 system, where R is the gas constant. While, only ~0.33$R\ln8$ of magnetic entropy is released below $T_N$, alongside the broad peak at $T_{sr}$~0.45 K, strong spin fluctuations are proposed to exist at low temperatures reflecting its 1D spin chain nature.

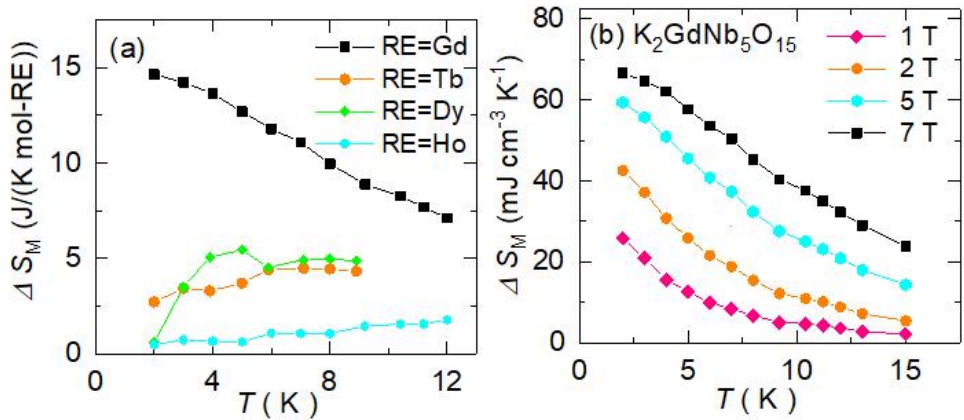

**Figure 9.** (a) Temperature dependence of magnetic entropy change (-$\Delta S_m$) of $K_2RENb_5O_{15}$ (RE= Gd, Tb, Dy, Ho) for field change $\Delta B$ = 7T, (b) Temperature dependence of -$\Delta S_m$ under different $\Delta B$ for $K_2GdNb_5O_{15}$ obtained from magnetization data.

Under applied field of 1 T, long-range magnetic ordering temperature $T_N$ is suppressed and the broad peak $T_{sr}$ shifts to high temperatures. As increased $B ≥$ 2T, $C_M(T)$ curves exhibit the Schottky-like anomaly. From the $S_M(T)$ curves at different fields shown in Figure 10c, magnetic entropy change ($\Delta S_m$) are obtained under different field variation $\Delta B = B_f-B_i$, as depicted in Figure 10d using for the evaluation of MC effect. Under $\Delta B$ = 7T, the maximum MCE appears at temperatures below 2 K, this is distinct from the dense $Gd^{3+}$-based compounds exhibiting the optical MCE in temperature well above 2 K.[58,59] Moreover, the maximum value of $\Delta S_m$=10.8 J/K/(mol-Gd)=0.62$R\ln(2S+1)$ under $\Delta B$ = 2 T occurs at ~ 1 K, this can be attractive for realizing sub-Kelvin cooling due to the lower $T_N$ of $K_2GdNb_5O_{15}$. In an adiabatic demagnetizing process as guided by the arrows in Figure 10b, magnetic entropy is reduced by 13.3(1) J/K/(mol-Gd) when a field of 5 T is applied at $T$ = 2 K, and the material is supposed to cool down to the minimum temperature $T_{min}$~ 175 mK, this is comparable to $T_{min}$ ~170 mK for Mn(NH$_4$)$_2$(SO$_4$)$_2$·6H$_2$O (MAS)[60] and $T_{min}$~125 mK in triangular-lattice KBaGd(BO$_3$)$_2$ polycrystals[61] as well as $T_{min}$~135 mK in NaYbGeO$_4$ polycrystals.[62] Also, a large volumetric entropy density ($S_{GS}$/vol.) is crucial to obtain long holding time for a practical use as magnetic coolants. For K$_2$GdNb$_5$O$_{15}$, the $S_{GS}$/vol ~93.2 mJ/(K cm$^3$) is larger than 70 mJ/(K cm$^3$) of MAS[60], but it is half of the value of KBaGd(BO$_3$)$_2$ [62] due to the very large interchain distances of Gd ions.

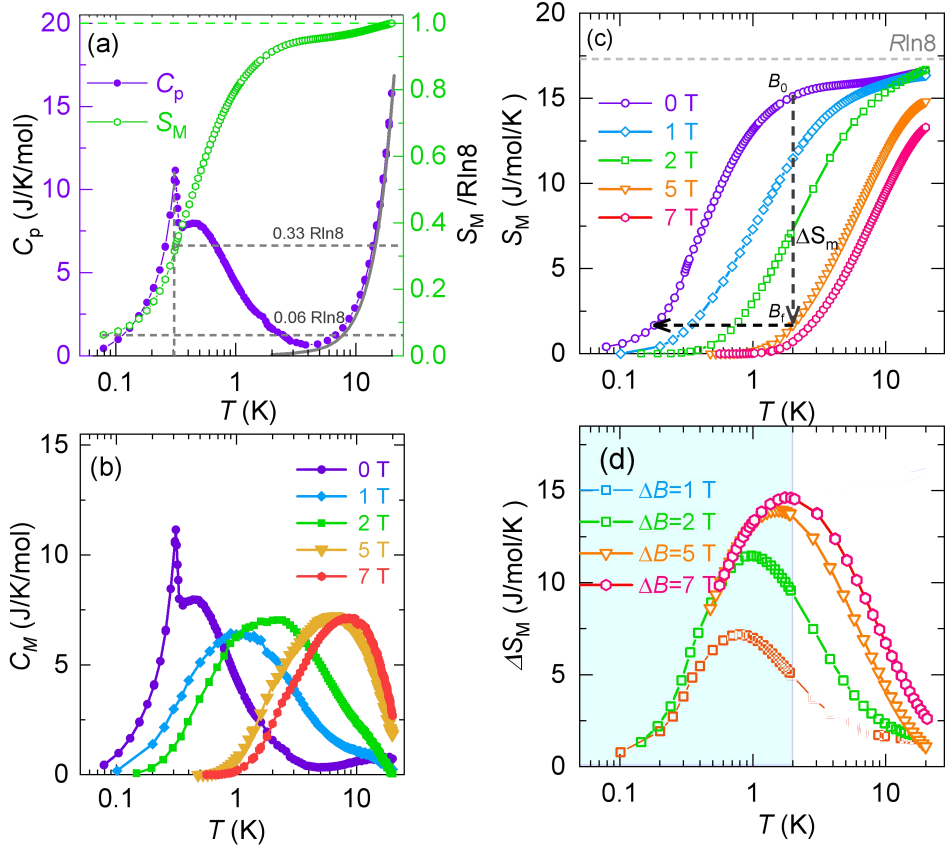

**Figure. 10.** (a) The zero-field specific heat $C_p$(T) and magnetic entropy $S_M$(T) curves of K$_2$GdNb$_5$O$_{15}$, the grey lines show the zero-field $C_p$(T) of nonmagnetic analog K$_2$LaNb$_5$O$_{15}$. (b) Temperature-dependence of magnetic specific heat $C_M$(T) of K$_2$GdNb$_5$O$_{15}$ under different fields, the entropy changes employed for cooling by adiabatic demagnetization are indicated by arrows. (c) The $S_M$(T) curves of K$_2$GdNb$_5$O$_{15}$ at different fields, the arrows show the adiabatic demagnetization process (d) Temperature dependence of -Δ$S_m$ at different Δ$B$.

Finally, we would like to highlight the influence of 1D spin chain structure on its magnetic ordering temperature of K$_2$GdNb$_5$O$_{15}$ by comparing with other Gd-based oxides reported in the literature. In the Gd-based antiferromagnets, both super-exchange and dipolar interactions between Gd$^{3+}$ ions are crucial on determining the spin state and depend on the distance of nearest neighbour (NN) $Gd^{3+}$ ions ($d_{NN}$), and the later one decay with $1/d_{NN}^3$, thus we plot the ordering temperature ($T_o$) versus $d_{NN}$ of $Gd^{3+}$ ions as shown in Figure 11. From that, we can identify that $K_2GdNb_5O_{15}$ exhibits the lowest $T_o$ among the insulating oxides with $d_{NN}$ of 3.5-4.0 Å, namely, the ones with comparable dipolar interactions. Here, the $T_o$~0.31 K is even lower than $T_o$~0.54 K of $Ca_2GdSbO_6$ with $d_{NN}$~5.873 Å and slightly higher than $T_o$~0.26 K of triangular-lattice $KBaGd(BO_3)_2$ with $d_{NN}$~5.46 Å,[62,63] while dipolar interactions for the latter two compounds only have 29-36% of that of $K_2GdNb_5O_{15}$. This fact highlights the importance of low dimensionality on reducing $T_O$ due to the strong spin fluctuation. Another important issue is that $K_2GdNb_5O_{15}$ and other family members of $K_2RENb_5O_{15}$ can exhibit ferroelectric behavior with large ferromagnetic polarization of 20-30 $\mu C/cm^2$ at room temperature. Thus, $K_2GdNb_5O_{15}$ can be considered as a multiferroic compounds with coexistent antiferromagnetic and ferroelectric order below $T_N$~0.31 K, the tunability of emergent magnetic behaviors are expected to be realized by applying electric field or strain field.

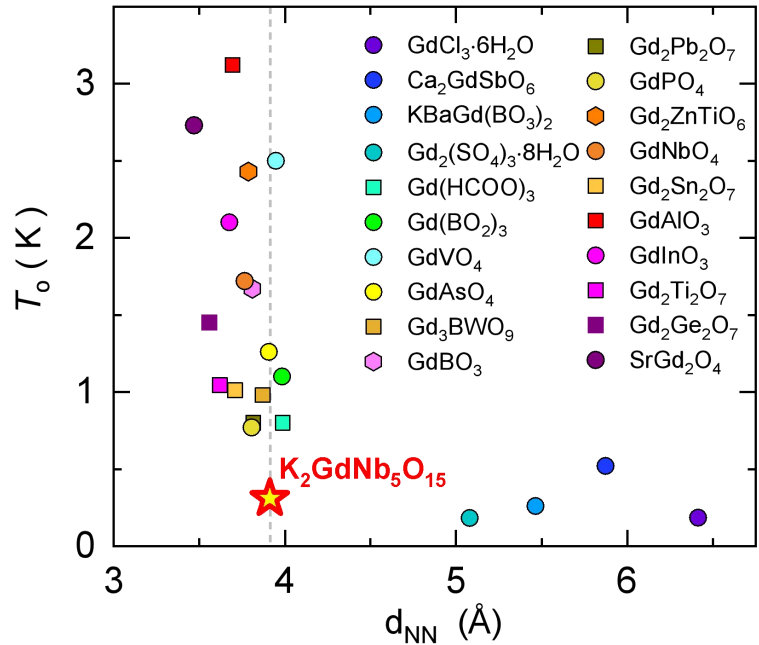

**Figure 11.** The ordering temperature ($T_o$) versus the distance of nearest-neighbor Gd ions ($d_{NN}$) for some of Gd-based oxides,[7,20,56,57, 61,63-79] the dashed line denotes the position of $d_{NN}$~3.91 Å of $K_2GdNb_5O_{15}$.

## ◼ CONCLUSION

In conclusion, a family of well-separated 1D spin chain compounds $K_2RENb_5O_{15}$ with tetragonal TTB structure are synthesized and magnetically characterized. It consists of face-sharing $REO_{12}$ tetrakaidecahedron chains along *c*-axis with well spatial separation of interchain distance $d_{inter}$ ~ 8.80-8.88 Å. No long-range magnetic order is detected down to 1.8 K for all family members, and different types of magnetic interactions are indicated for $K_2RENb_5O_{15}$ with different RE ions based on the low-*T* Curie-Weiss fitting. Among them, $K_2GdNb_5O_{15}$ with spin only magnetic moment ($S$=7/2), exhibits a long-range magnetic order with $T_N$~0.31 K and this ordering temperature is much lower than the ones of other Gd-based oxides with comparable NN distance between $Gd^{3+}$ ions. Alongside the low temperature specific heat results down to 80 mK, strong spin fluctuations are proposed to exist reflecting the low-dimension characteristics of 1D structure. The low-temperature magnetic entropy analysis on $K_2GdNb_5O_{15}$ reveal a large low-field ($\Delta B$ = 2 T) MC effect with maximum value $\Delta S_m$=10.8 J/K/(mol-Gd) realized at temperatures below 1 K, letting it attractive for realizing sub-Kelvin cooling. Moreover, $K_2RENb_5O_{15}$ provide a rare 1D spin chain system with the coexisting magnetic and ferroelectric order, that is promising to carry out the synthetic tunability on emergent magnetic behaviors in low dimensional system by combining the electric and magnetic fields.

## ■ ASSOCIATED CONTENT

Supporting Information: The Crystal Structure Parameters, differential scanning calorimetry (DSC) and X-band electron spin resonance (ESR) data of $K_2RENb_5O_{15}$ compounds. (PDF)

## AUTHOR INFORMATION

### Corresponding Author

**Liusuo Wu –**Department of Physics, Southern University of Science and Technology, Shenzhen, 518055, China; Shenzhen Key Laboratory of Advanced Quantum Functional Materials and Devices, Southern University of Science and Technology, Shenzhen, 518055, China; Email: wuls@sustech.edu.cn

**Zhaoming Tian –**Wuhan National High Magnetic Field Center and School of Physics, Huazhong University of Science and Technology, Wuhan, 430074, P. R. China*;* orcid.org/0000-0001-6538-3311; Email:tianzhaoming@hust.edu.cn

### Authors

**Qingyuan Zeng –**Wuhan National High Magnetic Field Center and School of Physics, Huazhong University of Science and Technology, Wuhan, 430074, P. R. China;

**Han Ge–**Department of Physics, Southern University Science and Technology, Shenzhen, 518055,

**Maofeng Wu–**Wuhan National High Magnetic Field Center and School of Physics, Huazhong University of Science and Technology, Wuhan, 430074, P. R. China;

**Shaoheng Ruan–**Wuhan National High Magnetic Field Center and School of Physics, Huazhong University of Science and Technology, Wuhan, 430074, P. R. China;

**Tiantian Li–**Department of Physics, Southern University Science and Technology, Shenzhen, 518055, China;

**Zhaosheng Wang–**Anhui Province Key Laboratory of Condensed Matter Physics at Extreme Conditions, High Magnetic Field Laboratory, Chinese Academy of Sciences, Hefei, 230031, China

**Jingxin Li–**Anhui Province Key Laboratory of Condensed Matter Physics at Extreme Conditions, High Magnetic Field Laboratory, Chinese Academy of Sciences, Hefei, 230031, China

**Langsheng Ling–**Anhui Province Key Laboratory of Condensed Matter Physics at Extreme Conditions, High Magnetic Field Laboratory, Chinese Academy of Sciences, Hefei, 230031, China

**Wei Tong–**Anhui Province Key Laboratory of Condensed Matter Physics at Extreme Conditions, High Magnetic Field Laboratory, Chinese Academy of Sciences, Hefei, 230031,China

**Shuai Huang–**Institute of Material Physics, Hangzhou Dianzi University, Key Laboratory of Novel Materials for Sensor of Zhejiang Province, Hangzhou, Zhejiang 310018, China.

**Andi Liu-** Wuhan National High Magnetic Field Center and School of Physics, Huazhong University of Science and Technology, Wuhan, 430074, P. R. China.

**Jin Zhou**- Wuhan National High Magnetic Field Center and School of Physics, Huazhong University of Science and Technology, Wuhan, 430074, P. R. China.

**Zhengcai Xia**- Wuhan National High Magnetic Field Center and School of Physics, Huazhong University of Science and Technology, Wuhan, 430074, P. R. China.

**Jieming Sheng–**Department of Physics, Southern University Science and Technology, Shenzhen, 518055, China.

**Author contributions**

Q. Zeng and H. Ge contribute equally to this work.

**Notes**

The authors declare no competing financial interest.

## ACKNOWLEDGEMENTS

This work was supported by the National Natural Science Foundation of China (Grant No. 11874158), the Fundamental Research Funds of Guangdong Province (Grant No. 2022A1515010658) and Guangdong Basic and Applied Basic Research Foundation (Grant No.2022B1515120020). A portion of this work was carried out at the Synergetic Extreme Condition User Facility (SECUF) by the synergetic extreme condition user facility (SECUF), and a portion of magnetic measurement was performed on the Steady High Magnetic Field Facilities, High Magnetic Field Laboratory. We would like to thank Liuting Gu and Hui fang for his assistance on the dielectric measurement and thank the staff of the analysis center of Huazhong University of Science and Technology for their assistance in structural characterizations.